# $P_{CN}$ calculations for Z=111 to Z=118

W. Loveland[1] and Liangyu Yao[1]

[1] Oregon State University, Corvallis, OR 97331 USA

**Abstract.** In previous publications [1,2] we presented evidence for the importance of spin in determining capture and evaporation residue cross sections in the synthesis of heavy nuclei. We extend the previous calculations which dealt with nuclei where $Z_{CN} \leq 110$ to the region of $Z_{CN}$ =111-118. We deduce a new systematics of the fusion probability $P_{CN}$ for these reactions.

**Keywords:** $P_{CN}$, superheavy element synthesis, spin mediated capture and evaporation residue cross sections.

## 1   Introduction

The cross section for producing a heavy evaporation residue in a complete fusion reaction can be written as a non-separable product of three factors, which express the capture cross section, the fusion probability and the survival probability.

$$\sigma_{EVR} = \frac{\pi h^2}{2\mu E} \sum_{\ell=0} (2\ell + 1) T(E, \ell) P_{CN} W_{sur}(E, \ell) \qquad (1)$$

Each of these factors is dependent on the spin, but the survival probability, $W_{sur}$, is zero or very small for higher spin values, effectively limiting the capture and fusion terms. Many partial waves contribute to the capture cross sections, but the higher partial waves result in non-surviving events. In this work, we examine the impact of restrictions on spin placed by the survival probabilities for compound nuclear reactions resulting in the synthesis of superheavy nuclei with $Z_{CN}$=111-118. In doing so, we extend the previous work [1,2] to treat the synthesis of the heaviest nuclei with $Z_{CN}$=111-118.

## 2   Methodology

As explained in [1], the formalism for calculating the survival, against fission, of a highly excited nucleus is relatively well-understood [3]. One starts with a single particle model [4] of the level density in which one allows the level density parameter to be a function of the excitation energy. Masses and shell corrections are taken from [5]. The deformation dependent collective enhancement of the level density is taken from [6]. The decay widths for decay by neutron, charged particle and γ-emission are calculated with standard formulas. Corrections for Kramers effects [7] are made to the fission widths. The fission barrier heights are calculated using liquid drop barriers and excitation energy dependent shell corrections.

We begin with the compilation of Duellmann of evaluated evaporation residue cross sections for reactions that produce nuclei with $Z_{CN}$ =111-118 [8]. For each reaction (projectile, target and beam energy), we calculated the spin dependent evaporation residue cross section assuming $P_{CN}$=1 using the "Empirical Model" of [3]. In [1], we presented evidence that this procedure results in a reasonable agreement between the calculated and measured spin dependence of the evaporation residue formation cross sections for the test case of $^{176}$Yb($^{48}$Ca,4n) $^{220}$Th reaction and for the $^{48}$Ca + $^{208}$Pb reaction. Loveland [9] has made a detailed examination of the strengths and weaknesses of models such as [3] and placed limits on how well these models work.

## 3. Results

There are 28 cases we have examined. A summary of the measured and calculated evaporation residue cross sections is given in Table 1. The fusion probability, $P_{CN}$, is taken as the ratios of the calculated to the measured evaporation residue cross sections since we have assumed $P_{CN}$ =1 in our calculations. As expected, the $P_{CN}$ values for the "cold fusion" reactions (1 n out) are orders of magnitude smaller than those for the hot fusion (2n-4n out) reactions. The deduced values of $P_{CN}$ generally get smaller as the product of the atomic numbers of the colliding nuclei, $Z_1Z_2$, increase.

**Table 1.** Measured and calculated evaporation residue cross sections for $Z_{CN}$ = 111-118

| Beam | Target | Channel | $\sigma_{meas}$(pb) | $\sigma_{calc}$(pb) | $P_{CN}$ | Ref |
|---|---|---|---|---|---|---|
| $^{64}$Ni | $^{209}$Bi | 1n | $3.5^{+1.9}_{-1.3}$ | 6910 | 0.000507 | 10 |
| $^{65}$Cu | $^{208}$Pb | 1n | $1.7^{+3.9}_{-1.4}$ | 20500 | 8.3e-05 | 11 |
| $^{48}$Ca | $^{238}$U | 3n | $2.5^{+1.8}_{-1.1}$ | 60 | 0.0417 | 12 |
| $^{48}$Ca | $^{238}$U | 4n | $0.7^{+0.6}_{-0.3}$ | 425 | 0.00169 | 13 |
| $^{48}$Ca | $^{238}$U | 4n | $0.6^{+1.6}_{-0.5}$ | 7 | 0.0857 | 12 |
| $^{70}$Zn | $^{208}$Pb | 1n | $0.5^{+1.1}_{-0.4}$ | 5e+06 | 1e-07 | 14 |
| $^{48}$Ca | $^{237}$Np | 3n | 0.9 | 5 | 0.18 | 15 |
| $^{70}$Zn | $^{209}$Bi | 1n | $0.022^{+0.020}_{-0.013}$ | 940000 | 2.34e-08 | 16 |
| $^{48}$Ca | $^{239}$Pu | 3n | 0.23 | 16 | 0.0144 | 17 |
| $^{48}$Ca | $^{240}$Pu | 3n | $2.5^{+2.9}_{-1.4}$ | 62 | 0.0403 | 17 |
| $^{48}$Ca | $^{242}$Pu | 2n | 0.5 | 244 | 0.00205 | |
| $^{48}$Ca | $^{242}$Pu | 3n | $3.6^{+3.4}_{-1.7}$ | 78 | 0.0463 | 12 |
| $^{48}$Ca | $^{242}$Pu | 4n | $4.5^{+3.6}_{-1.9}$ | 129 | 0.0349 | 12 |
| $^{48}$Ca | $^{242}$Pu | 5n | $0.6^{+0.9}_{-0.5}$ | 11.6 | 0.0517 | 18 |
| $^{48}$Ca | $^{244}$Pu | 3n | $8^{+7.4}_{-4.5}$ | 180 | 0.0444 | 19 |
| $^{48}$Ca | $^{244}$Pu | 4n | $9.8^{+3.9}_{-3.1}$ | 220 | 0.0445 | 19 |
| $^{48}$Ca | $^{244}$Pu | 5n | $1.1^{+2.6}_{-0.9}$ | 9.2 | 0.120 | 20 |
| $^{48}$Ca | $^{243}$Am | 2n | $2.5^{+2.7}_{-1.5}$ | 15.4 | 0.162 | 21 |
| $^{48}$Ca | $^{243}$Am | 3n | $8.5^{+6.4}_{-3.7}$ | 660 | 0.0129 | 22 |
| $^{48}$Ca | $^{243}$Am | 4n | $0.9^{+3.2}_{-0.8}$ | 169 | 0.00533 | 23 |
| $^{48}$Ca | $^{245}$Cm | 2n | 0.9 | 6.89 | 0.131 | 20 |
| $^{48}$Ca | $^{245}$Cm | 3n | $3.7^{+3.6}_{-1.8}$ | 229 | 0.0162 | 24 |
| $^{48}$Ca | $^{245}$Cm | 4n | 0.8 | 95 | 0.00842 | 24 |
| $^{48}$Ca | $^{248}$Cm | 3n | 1.2 | 166 | 0.00723 | 12 |
| $^{48}$Ca | $^{248}$Cm | 4n | 3.4 | 652 | 0.00522 | 25 |
| $^{48}$Ca | $^{249}$Bk | 3n | $1.1^{+1.2}_{-0.6}$ | 1660 | 0.000663 | 26 |
| $^{48}$Ca | $^{249}$Bk | 4n | $2.4^{+3.3}_{-1.4}$ | 333 | 0.00721 | 26 |
| $^{48}$Ca | $^{249}$Cf | 2n | 0.9 | 50.9 | 0.0177 | 24 |

In Figure 1, we show the $P_{CN}$ values, sorted by exit channel for the hot fusion reactions, as a function of the simple scaling variable, $Z_1Z_2$, the product of the atomic numbers of the reacting nuclei.

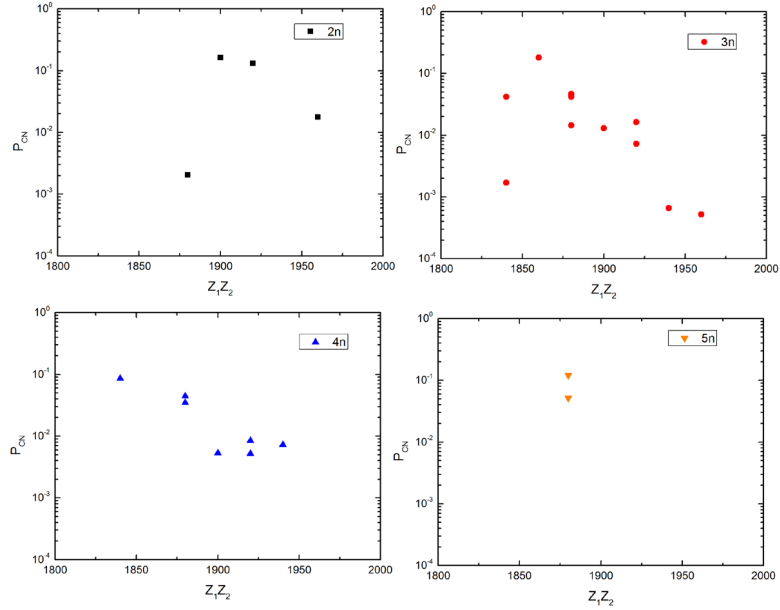

Figure 1. The calculated values of $P_{CN}$ for various exit channels as a function of the scaling variable $Z_1Z_2$.

The use of other scaling variables, such as $x_{CN}$, $x_{eff}$ and $x_m$ does not significantly improve the description of the data. $x_{CN}$ is defined as

$$x_{CN} = \frac{Z_{CN}^2/A_{CN}}{50.883(1 - 1.7826(\frac{A_{CN} - 2Z_{CN}}{A_{CN}})^2)} \quad (2)$$

$x_{eff}$ is defined as

$$x_{eff} = \frac{\frac{4Z_P Z_T}{A_P^{\frac{1}{3}} A_T^{\frac{1}{3}} \left( A_P^{\frac{1}{3}} + A_T^{\frac{1}{3}} \right)}}{50.883(1 - 1.7826( \left(\frac{A_{CN} - 2Z_{CN}}{A_{CN}}\right)^2}$$

$x_m$ is defined as

$$x_m = 0.25 x_{CN} + 0.75 x_{eff}$$

All of these scaling variables seek to relate $P_{CN}$ to the balance of attractive and repulsive forces in the reaction entrance channel. Clearly there is a certain amount of "spatter" in the plots of $P_{CN}$ vs. $Z_1 Z_2$. In part, this "spatter" is due to the uncertainties in the measured evaporation residue cross sections which are typically uncertain to the measured value. (Loveland [9] has shown that these uncertainties in $P_{CN}$ can lead to order of magnitude uncertainties in estimations of the production cross sections for elements 119 and 120, challenging experimentalists dealing with fb production cross sections.)

If we use the simple $Z_1 Z_2$ scaling factor for the 3n and 4n reactions, then we can write a simple formula for the 3n channel as $P_{CN}(3n) = -0.019 Z_1 Z_2 + 35.0$ and for the 4n channel $P_{CN}(4n) = -0.013 Z_1 Z_2 + 23.2$.

We can ask how well these new values of $P_{CN}$ agree with previous measurements and theoretical predictions. Kozulin et al. [27] have reported measurements of $P_{CN}$ based upon mass-energy distributions of fission-like fragments from a variety of reactions. In Figure 2, we compare our values of $P_{CN}$ with the Kozulin et al. measurements. Given the intrinsic large uncertainties in our deduced $P_{CN}$ values, the agreement between the measurements seems satisfactory.

How do our measured values of $P_{CN}$ compare with various theoretical predictions of $P_{CN}$? Given our methodology, there is no surprise that our deduced values of $P_{CN}$ agree well with the predictions of Zagrebaev [29]. How about other predictions? In Figure 3, we compare our deduced values of $P_{CN}$ with predictions of Nasirov et al.[28]. For the hot fusion reactions ($Z_1 Z_2 = 1800$-$2000$), the agreement seems reasonable but there is a stark disagreement for the cold fusion cases.

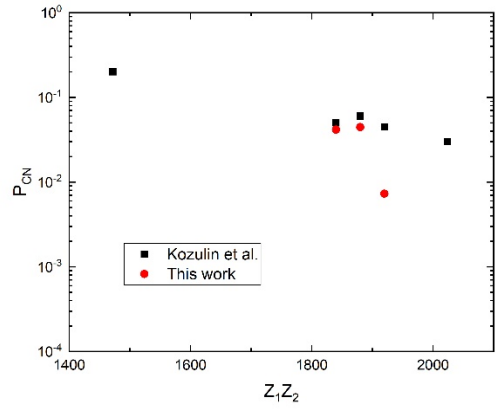

Figure 2. Comparison of the measurements of $P_{CN}$ in this work with that of [27].

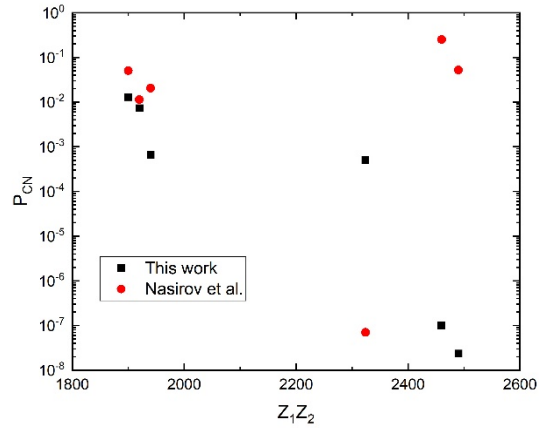

Figure 3. Comparison of our measured values of $P_{CN}$ with the predictions of [28]

## 4. Conclusions

What have we learned from this study? We have extended the systematics of $P_{CN}$ to cases involving the synthesis of elements 111-118. We have parameterized the new values of $P_{CN}$ with a simple linear fit that might be useful in predictions of cross sections for the synthesis of elements 119 and 120. We have compared our measurements with previous measurements and theoretical predictions.

## 5. Acknowledgments

This work was supported in part by the U.S. Department of Energy, Office of Science, Office of Nuclear Physics under Grant No. DE-SC0014380.